# Incorporating Surrogate Information for Adaptive Subgroup Enrichment Design with Sample Size Re-estimation

Liwen Wu[a], Qing Li[a*], Mengya Liu[a] and Jianchang Lin[a]

[a]*Takeda Pharmaceuticals, Cambridge, MA, USA*

*corresponding email: qing.li2@taketa.com

# Incorporating Surrogate Information for Adaptive Subgroup Enrichment Design with Sample Size Re-estimation

Adaptive subgroup enrichment design is an efficient design framework that allows accelerated development for investigational treatments while also having flexibility in population selection within the course of the trial. The adaptive decision at the interim analysis is commonly made based on the conditional probability of trial success. However, one of the critical challenges for such adaptive designs is immature data for interim decisions, particularly in the targeted subgroup with a limited sample size at the first stage of the trial. In this paper, we improve the interim decision making by incorporating information from surrogate endpoints when estimating conditional power at the interim analysis, by predicting the primary treatment effect based on the observed surrogate endpoint and prior knowledge or historical data about the relationship between endpoints. Modified conditional power is developed for both selecting the patient population to be enrolled after the interim analysis and sample size re-estimation. In the simulation study, our proposed design shows a higher chance to make desirable interim decisions and achieves higher overall power, while controlling the overall type I error. This performance is robust over drift of prior knowledge from the true relationship between two endpoints. We also demonstrate the application of our proposed design in two case studies in oncology and vaccine trials.

**Keywords:** Adaptive subgroup enrichment design; Surrogate endpoint; Sample size re-estimation; Modified conditional power

# 1. Introduction

As drug developers aim to develop effective drugs for patients with unmet medical needs, sponsors usually attempt to develop drugs in faster and more efficient ways. Adaptive design, which allows prospectively planned modification to one or more aspects of the design based on accumulating data from subjects in the trial (FDA, 2019), becomes critical in the drug development process. In particular, phase 2/3 seamless design is one of such adaptive designs that can accelerate the drug development process by minimizing the white space between phase 2 and phase 3 studies and potentially reducing the sample size by combing the phase 2 data into the final analysis after phase 3 (Li, Lin & Lin, 2020). A typical seamless phase 2/3 design consists of two stages: a learning stage (or first stage), corresponding to phase 2 under the traditional trial framework, where data and knowledge about the investigational treatments are accumulated to make a suitable adaptation for the trial at the interim analysis; followed by a confirmatory stage (or second stage) which represents the traditional phase 3, or the confirmatory trial. Some widely used adaptive frameworks with different interim decisions have been established in the past decades. One example is sample size re-estimation (SSR), where sample size increasing is allowed for the second stage of the trial based on interim decisions (Chen, DeMets, & Gordon, 2004; Cui, Hung, & Wang, 1999; Gao, Ware, & Mehta, 2008). Another example is subgroup enrichment, which limits the future recruitment to a specific subgroup(s) that demonstrate promising treatment effects at interim analyses (Alosh et al., 2015; Friede, Stallard, & Parsons, 2020; Lin, Bunn, & Liu, 2019; Magnusson & Turnbull, 2013; Sinha et al., 2019; Stallard, Hamborg, Parsons, & Friede, 2014). For these types of adaptive design, conditional power, ie. the probability of trial to succeed by the end of the trial based on interim data, is commonly used to facilitate interim decision making (Gao et al., 2008; Mehta & Pocock, 2011), while many other studies employ the approach of group sequential framework with the joint distribution of the test statistics to obtain the decision boundaries (Friede et al., 2020; Jennison & Turnbull, 1999; Magnusson & Turnbull, 2013). The adaptive seamless 2/3 design with both SSR and subgroup enrichment features is illustrated in Figure 1.

[Figure 1]

A critical challenge in designing an adaptive seamless trial is the timing of the interim analysis. Interim analysis designed later in the trial could take advantage of more mature data and share similar statistical properties of final analysis in a non-adaptive trial. On the other hand, late interim analysis reduces trial flexibility for adaptation, since recruitment and randomization could be nearly completed. In some therapeutic areas like oncology, vaccine and chronic diseases, where the primary survival endpoint such as Overall Survival (OS) and Progression Free Survival (PFS) may take longer time to observe, it is appealing to take into consideration a short-term surrogate endpoint at the interim analysis, if approved to be a good proxy of the primary endpoint. As shown in Figure 2, a good surrogate endpoint candidate is expected to mature earlier than the primary endpoint, thus more surrogate endpoint data will be available at the interim analysis. For example, in newly diagnosed Philadelphia Chromosome–Positive Acute Lymphoblastic Leukemia (ND_Ph+ ALL), Minimal Residual Disease (MRD)-negative Complete Response (CR) is often used as a surrogate endpoint for Event-Free-Survival (EFS), so it avoids potentially treatment switching confounding effects when analyzing survival data (Berry et al., 2017). Similarly, for multiple myeloma diesease, the post-induction CR rates and Minimal Residual Disease (MRD) are identified as important predictors for PFS and OS (Harousseau et al., 2010; Munshi et al., 2017). This prior knowledge of surrogacy has the great potential to contribute to informed decision making at the interim analysis.

[Figure 2]

In comparison, an earlier interim analysis allows flexible adaptation, but it may subject to bias introduced by immature data or delayed treatment effects. For example, a long-term primary endpoint of clinical remission at week 52 was used for both interim sample-size adaptation and final analysis for VARSITY study, a head-to-head study of Vedolizumab versus Adalimumab (Sands et al., 2019). If using only data available at week 52 for interim decision making, the extensive clinical remission data before week 52 from subjects who are not included in the interim analysis were overlooked. In the example of EMERGE & ENGAGE studies for Aducanumab, Clinical Dementia Rating-Sum of Boxes (CDR-SB) at 18 months was the primary efficacy endpoint to be considered for interim decision (ClinicalTrials.gov, 2019). This long-term endpoint was only available for a subset of subjects who had the opportunity to complete (OTC) by the time of the interim analysis and the study was stopped for futility. However, it was subsequently revealed in an analysis after a longer period of follow up that aducanumab reduced clinical decline in patients with early Alzheimer's disease as measured by the same endpoint later. This led to a reversal of the sponsor's decision to terminate the development activity of the compound. A possible explanation is that the interim futility decision was based on inadequate data of the long-term endpoint. This example further underscores the value of incorporating available auxiliary information from patients who may not have had OTC at the time of the interim, to facilitate "better-informed" interim decision-making.

The primary hurdle of including surrogate endpoints in trial design is to find good surrogate endpoints and to maintain statistical integrity with this additional information. Prentice (1989) firstly established a set of formal criteria for surrogate endpoints in statistical language and outlined a validation procedure. Many studies following this route investigated the possibility of substitute the primary endpoints by surrogates and how to model the data from after the substitution (Buyse & Molenberghs, 1998; Buyse, Molenberghs, Burzykowski,

Renard, and Geys, 2000; Freedman, Graubard, and Schatzkin, 1992). Jenkins, Stone, and Jennison (2011) used an example of applying surrogate endpoints at the interim analysis in a seamless phase 2/3 trial with subgroup enrichment. Chen et al. (2018) discussed a new design with a more flexible specification of endpoints at each stage without explicit adjustment to overall type I error, as long as the endpoints follow certain correlation constraint. Fleming, Prentice, Pepe, and Glidden (1994) explored another approach where they considered the surrogate endpoint as 'auxiliary variable' and include its information into an augmented likelihood estimator together with the primary endpoint. Li, Lin, Liu, Wu, and Liu (2021) incorporated surrogate endpoints in a weighted function when estimating the conditional power and demonstrated its benefits in catching delayed treatment effects. This approach, instead of replacing the primary endpoint completely, requires a weaker assumption on the correlation structure between the primary and surrogate endpoints.

In this paper, we propose a new adaptive seamless phase 2/3 subgroup enrichment design that incorporates surrogate information at interim decision making while allowing sample size re-estimation (SSR). In Section 2, we will present the notation and statistical methodology of the proposed design. The operating characteristics will be examined via a simulation study in Section 3. Section 4 will demonstrate two case studies, in oncology and vaccine respectively, to demonstrate the practical usage of our proposed method in clinical trials. We will conclude this paper with discussion and conclusion in Section 5.

## 2. Methods

### *2.1 Notation in adaptive subgroup enrichment trial*

Suppose in a two-stage adaptive subgroup enrichment trial, the full target population $F$ can be characterized by two mutually exclusive subgroups, a target subgroup $S$, which is proportion $\tau$ of the full population, and the complementary subgroup $S^c$. For simplicity, assume the primary endpoint of a trial, denoted by $Y$, for the full population follows Gaussian distribution $N\left(\mu_k^{(F)}, \left(\sigma_k^{(F)}\right)^2\right)$, and within subgroup $S$ follows $N\left(\mu_k^{(S)}, \left(\sigma_k^{(S)}\right)^2\right)$, for the two treatments $k = 0,1$. At each stage $j$, the estimators of the mean treatment effects for the target subgroup and full population naturally follow a bivariate gaussian distribution

$$\begin{pmatrix} \hat{\mu}_{kj}^{(S)} \\ \hat{\mu}_{kj}^{(F)} \end{pmatrix} \sim N\left( \begin{pmatrix} \mu_k^{(S)} \\ \mu_k^{(F)} \end{pmatrix}, \begin{pmatrix} \frac{\left(\sigma_k^{(S)}\right)^2}{\tau n_{kj}} & \frac{\left(\sigma_k^{(F)}\right)^2}{n_{kj}} \\ \frac{\left(\sigma_k^{(F)}\right)^2}{n_{kj}} & \frac{\left(\sigma_k^{(F)}\right)^2}{n_{kj}} \end{pmatrix} \right), \quad k = 0,1; j = 1,2,$$

where $n_{kj}$ denotes the cumulative sample size that is allocated to treatment $k$ at stage $j$, $n_j = n_{1j} + n_{0j}$. While across two stages $cov\left(\hat{\mu}_{kj}^{(S)}, \hat{\mu}_{kj'}^{(F)}\right) = \frac{\left(\sigma_k^{(F)}\right)^2}{n_{k,max(j,j')}}$, for $j \neq j'$.

At each stage, the two elementary null hypotheses could be formulated as:
$H_{0j}^{(S)}: \mu_1^{(S)} = \mu_0^{(S)}$ or $H_{0j}^{(F)}: \mu_1^{(F)} = \mu_0^{(F)}$, to test any between treatment differences within the subgroup or the full population. The test statistic for each elementary null hypothesis is defined as:

$$Z_j^{(s)} = \frac{\hat{\mu}_{1j}^{(s)} - \hat{\mu}_{0j}^{(s)}}{\sqrt{\frac{\left(\sigma_1^{(s)}\right)^2}{n_{1j}^{(s)}} + \frac{\left(\sigma_0^{(s)}\right)^2}{n_{0j}^{(s)}}}}, \quad j = 1,2; s = S, F.$$

Under each null hypothesis, $Z_j^{(s)} \sim N(0,1)$.

With the independent increment assumption, it can be shown that $cov\left(Z_1^{(S)}, Z_2^{(S)}\right) = cov\left(Z_1^{(F)}, Z_2^{(F)}\right) = \sqrt{n_1/n_2}$. Let $\sigma_j^{(s)\, pool} = \sqrt{\left(\sigma_1^{(s)}\right)^2/n_{1j}^{(s)} + \left(\sigma_0^{(s)}\right)^2/n_{0j}^{(s)}}$ for $j = 1,2$ and $s = S, F$. After some algebraic steps,

$$cov\left(Z_1^{(S)}, Z_1^{(F)}\right) = \left(\sigma_1^{(S)\,pool}\sigma_1^{(F)\,pool}\right)^{-1}\left(\frac{\left(\sigma_0^{(F)}\right)^2}{n_{01}} + \frac{\left(\sigma_1^{(F)}\right)^2}{n_{11}}\right), \text{ and}$$

$$cov\left(Z_1^{(S)}, Z_2^{(F)}\right) = \left(\sigma_1^{(S)\,pool}\sigma_2^{(F)\,pool}\right)^{-1}\left(\frac{\left(\sigma_0^{(F)}\right)^2}{n_{02}} + \frac{\left(\sigma_1^{(F)}\right)^2}{n_{12}}\right).$$

In a subgroup enrichment trial, one goal for the interim analysis is to decide whether to include the full population or only a target subgroup in the next stage of the trial. Stallard et al. (2014) discussed statistical considerations in this adaptation and compared different methods for population selection. In this paper, the focus is on the conditional power (CP) based approach. Based on the interim readouts and the incremental sample size in the second stage ($\tilde{n}_2 = n_2 - n_1$), the CP is defined as

$$CP_s\left(\hat{Z}_1^{(s)}, \tilde{n}_2\right) = \Pr\left(Z_2^{(s)} > Z_{1-\alpha}\middle|\hat{Z}_1^{(s)}, \tilde{n}_2\right)$$

$$= 1 - \Phi\left(\frac{Z_\alpha\sqrt{n_2} - \hat{Z}_1^{(s)}\sqrt{n_1}}{\sqrt{\tilde{n}_2}} - \hat{Z}_1^{(s)}\sqrt{\frac{\tilde{n}_2}{n_1}}\right), \qquad (1)$$

for $s = S, F$; where $\alpha$ is the Type-I error rate.

Mehta and Pocock (2011) outlined a general framework for using CP to perform SSR at interim analysis. For each group $s = S, F$, if $CP_s\left(\hat{Z}_1^{(s)}, \tilde{n}_2\right) < 1 - \beta$, where $1 - \beta$ is the pre-specified power, the corresponding sample size can be increased in the next stage to achieve a desired overall power at the final analysis. The adaptive sample size $\tilde{n}_2^*$ can be obtained by solving $CP_s\left(\hat{Z}_1^{(s)}, \tilde{n}_2^*\right) = 1 - \beta$. Ideally, we could extend to a very large trial to preserve designed power if the observed CP is low, but it will lead to great inflation of the type I error. Thus, a minimum threshold $\delta^{(s)}$ would be needed to maintain our family-wise type I error in the 'promising zone'. Similar to Mehta, Liu, and Theuer (2019), one could define the interim decision zones with conditional power thresholds $\delta^{(S)}, \delta^{(F)}$ for both subgroup population and full population as in Table 1.

[Table 1]

*2.2 Modified conditional power with short-term endpoint*

One drawback of such subgroup enrichment adaptation is that when the interim analysis takes place early, the interim data might not be mature enough to make a reliable adaptation decision. Due to the limited sample size, subgroups are generally not sufficiently powered to detect a significant difference in treatment effects. Particularly, when the subgroup is a small proportion of the full population, the CP estimation may be biased (Lin et al., 2019). In light of this, we proposed modification on the conditional power to incorporate information from surrogate endpoints.

To incorporate surrogate information in CP calculation, we followed a similar framework proposed by Li et al. (2021) where surrogate endpoints are seen as predictors for the corresponding primary endpoints, and their relationship can be expressed through a prediction model, sometimes a simple linear model. Denote the surrogate endpoint as X, and the treatment effect for surrogate endpoint X as $\theta^{s,X}$ for the corresponding group of patients $s = S, F$. For specific endpoints to be considered in the design, Li et al. (2021) assumed that there is some prior knowledge that is readily available so that one can fit a linear regression function $f: \hat{\theta}^{s,X} \to \hat{Z}^{(s)}$, in the form $f(\hat{\theta}^{s,X}) = a + b * \hat{\theta}^{s,X}$, where $a$ and $b$ represents constant values to be estimated. This kind of prior information can be obtained from existing clinical trials, published data, real-world-data (RWD) or meta-analyses. An example is the strong association between trial-level PFS hazard ratio and ORR odds ratio found in patients with advanced NSCLC. Based on data collected from 15 trials, researchers build a linear regression model between the two endpoints in the trial-level analysis (Blumenthal et al.

2015). In this paper, we assume such prior knowledge, referred to as a 'historical model' hereafter in this paper, is known before applying the proposed adaptive enrichment design.

Supposing the test statistic $\hat{Z}_1^{(s)}$ for the primary endpoint and the treatment effect $\hat{\theta}^{s,X}$ for the surrogate endpoint are observed at the interim readouts, we can use the weighting method proposed by Li et al. (2021) to combine these information. Let $t = n_1/n_2$ represent the information fraction at the interim analysis, the modified CP can be calculated as following,

$$CP_s^{(1)}\left(\hat{Z}_1^{(s)}, \tilde{n}_2\right) = 1 - \Phi\left(\frac{Z_\alpha \sqrt{n_2} - \hat{Z}_1^{(s)}\sqrt{n_1}}{\sqrt{\tilde{n}_2}} - \frac{\sqrt{\tilde{n}_2}}{\sqrt{n_1}}\left(\hat{Z}_1^{(s)} t + f(\hat{\theta}^{s,X})(1-t)\right)\right), \quad (2)$$

$$CP_s^{(2)}\left(\hat{Z}_1^{(s)}, \tilde{n}_2\right) = 1 - \Phi\left(\frac{Z_\alpha \sqrt{n_2} - \hat{Z}_1^{(s)}\sqrt{n_1}}{\sqrt{\tilde{n}_2}} - \frac{\sqrt{\tilde{n}_2}}{\sqrt{n_1}}\left(\frac{\hat{Z}_1^{(s)} f(\hat{\theta}^{s,X})}{\hat{Z}_1^{(s)}(1-t) + f(\hat{\theta}^{s,X}) t}\right)\right), \quad (3)$$

$$CP_s^{(3)}\left(\hat{Z}_1^{(s)}, \tilde{n}_2\right) = 1 - \Phi\left(\frac{Z_\alpha \sqrt{n_2} - \hat{Z}_1^{(s)}\sqrt{n_1}}{\sqrt{\tilde{n}_2}} - \frac{\sqrt{\tilde{n}_2}}{\sqrt{n_1}}\left(\frac{\hat{Z}_1^{(s)} f(\hat{\theta}^{s,X})}{\hat{Z}_1^{(s)}(1-F_c(t)) + f(\hat{\theta}^{s,X}) F_c(t)}\right)\right), \quad (4)$$

for $s = S, F$, where $f(\hat{\theta}^{s,X})$ represent a predicted primary treatment effect by plugging the observed $\hat{\theta}^{s,X}$ in a known historical model, and $F_c(t)$ is the cumulative density function of the survival time for the control group at information fraction $t$. Equations (2)-(4) represent three different weighting strategies to incorporate the directly observed treatment effect and the predicted treatment effect based on the surrogate endpoint. Equation (2) can be seen as a weighted average of these two counterparts based on the information proportion, $t$, at the interim analysis, while Equation (3) and (4) are derived based on a semiparametric analysis of the primary and surrogate survival endpoints (Li et al., 2021; Yang & Prentice, 2005).

### 2.3 Interim decision rule for population selection

After obtaining interim readouts from $n_1$ patients in the first stage, the modified conditional power $CP_S^*\left(\hat{Z}_1^{(S)}, \tilde{n}_2\right)$ and $CP_F^*\left(\hat{Z}_1^{(F)}, \tilde{n}_2\right)$ could be calculated for both the subgroup and the full population, where $CP_S^*$ is a simplified notation for any of the three modified conditional power estimates aforementioned in Equations (2)-(4). Based on these calculated CP, the interim decision should be made based on a pre-specified decision tree as illustrated in Figure 3. If the CPs are in the Promising zone, SSR will be performed to extended $\tilde{n}_2^*$ such that $CP_F^*\left(\hat{Z}_1^{(F)}, \tilde{n}_2^*\right) = 1 - \beta$. In this scenario, the designed power is achieved, and recruitment for the full population will continue when proceeding into the second stage. At the end of the trial, the efficacy test will be based on the full population using the primary endpoint. If the conditional powers fall in the Enrichment zone, SSR will be performed for $\tilde{n}_2^*$ such that $CP_S^*\left(\hat{Z}_1^{(S)}, \tilde{n}_2^*\right) = 1 - \beta$ and second stage recruitment will be limited to the subgroup $S$ only. The final analysis will be based on the target subgroup accordingly. If the conditional powers fall into the Favorable or Unfavorable zones, recruitment will continue as planned without any sample size adaptation, and the test for efficacy in full population will be performed at the end of the trial. Futility is also allowed to terminate the trial early if no efficacy signal is observed in neither the full population nor the subgroup at the interim analysis.

[Figure 3]

### 2.4 Type I error control

Overall type I error control would be critical in such adaptive design due to multiple looking of the efficacy endpoints across the study. The combination test (CHW test) proposed by Cui et al. (1999) is used to maintain the overall type I error,

$$Z_{chw}^{(s)} = w_1 Z_1^{(s)} + w_2 \tilde{Z}_2^{(s)} \quad , s = S, F.$$

The weights $\left(w_1 = \sqrt{\frac{n_1^{(s)}}{n_2^{(s)}}}, w_2 = \sqrt{\frac{\tilde{n}_2^{(s)}}{n_2^{(s)}}}\right)$ are based on the planned sample size, or number of events if the design is event driven, in which it remains unchanged regardless of the adaptive sample size $\tilde{n}_2^*$ estimated with modified conditional power. This method is widely used in adaptive trials and proven to be robust over different types of endpoints. However, for time-to-event endpoints, since some patients enrolled in the first stage (i.e. cohort 1) may have events observed in the second stage, $Z_1^{(S)}$ and $\tilde{Z}_2^{(S)}$ are no longer independent. Jenkins et al. (2011) demonstrated that the CHW test was still valid as long as the number of events to observe before and after interim analysis in cohort 1 could be prespecified.

Another potential place where multiplicity may arise is the subgroup selection procedure. This type of multiplicity could be controlled locally by using the closure testing principle (Marcus, Eric, and Gabriel, 1976). Consider two elementary hypotheses $H_0^{(S)}: \mu_1^{(S)} = \mu_0^{(S)}$ or $H_0^{(F)}: \mu_1^{(F)} = \mu_0^{(F)}$, under the closure testing principle, any of them can be rejected if and only if the intersection of both hypotheses $H_0^{(FS)} = H_0^{(F)} \cap H_0^{(S)}$ and itself are rejected at the same time. For the stagewise data, Hochberg correction (Hochberg 1988 & Jenkins et al. 2011) is used with equal weights of $H_0^{(F)}$ and $H_0^{(S)}$, which leads to $Z_j^{(FS)} = \Phi^{-1}\left(\min\left(2\min\left\{\Phi\left(Z_j^{(F)}\right), \Phi\left(Z_j^{(S)}\right)\right\}, \max\left\{\Phi\left(Z_j^{(F)}\right), \Phi\left(Z_j^{(S)}\right)\right\}\right)\right), j = 1,2$ for a lower 1-sided test. Notice that $Z_j^{(FS)}$ may only be estimable for information obtained at the first stage before the interim analysis, where both $Z_1^{(F)}$ and $Z_1^{(S)}$ are estimated. If the interim decision is to perform subgroup enrichment, only patients belong to the subgroup would be enrolled in the second stage, thus $Z_2^{(F)}$ is not estimable. We follow a similar approach to Jenkins et al. 2011 to calculate the final test statistics given the population selected at the interim analysis:

- If full population is selected at the interim analysis, reject both

$$H_0^{(F)}: Z_{chw}^{(F)} = w_1 Z_1^{(F)} + w_2 \tilde{Z}_2^{(F)},$$

$$H_0^{(FS)}: Z_{chw}^{(FS)} = w_1 Z_1^{(FS)} + w_2 \tilde{Z}_2^{(F)}.$$

- If subgroup is selected at the interim analysis, test reject both

$$H_0^{(S)}: Z_{chw}^{(S)} = w_1 Z_1^{(S)} + w_2 \tilde{Z}_2^{(S)},$$

$$H_0^{(FS)}: Z_{chw}^{(FS)} = w_1 Z_1^{(FS)} + w_2 \tilde{Z}_2^{(S)}.$$

to show evidence supporting the alternative hypothesis for the test treatment in the population selected. Each elementary hypothesis is rejected when the CHW test statistic falls into the rejection area. For a lower one-sided test at $\alpha = 0.025$, we reject any of them if $Z_{chw}^{(\cdot)} < -1.96$.

*2.5 Selection of design parameters*

It is critical to evaluate the operating characteristics, such as power, average study duration, and average sample size to optimize the adaptive designs. Mehta et al. (2019) comprehensively summarized important design parameters in formulating such an adaptive enrichment trial. Selecting optimal parameters for sample sizes, the timing of interim analysis, and decision cutoffs requires combined knowledge of the statistical property and practical considerations to achieve desirable power while maintaining type I error, through extensive simulation. Here we present a sample study design as illustrated in Figure 4, which will be further assessed by simulation studies in Section 3.

[Figure 4]

Assuming the hazard ratio for overall survival in the full population and target subgroup is equally 0.6 under the alternative hypothesis, the number of events needed to achieve 90% power is about 160, with one-sided $\alpha = 0.025$. The number of patients recruited in the trial is calculated by a feasible recruitment and event rate. We timed our interim analysis after observing 40 events in the full population and will keep following up on the first cohort of patients until observing 60 events in total (i.e. the interim analysis is planned at 2/3 information fraction). In the second stage, we enroll 200 more patients, and follow them until observing 100 events or the event size re-estimated by the steps introduced earlier in this section. The interim decision cutoffs can be selected by evaluating their operating characteristics under different hypotheses. We will present the operating characteristics of this example design in the following section.

## 3. Simulation Study

### 3.1 Simulation setup

In this simulation study, we considered a seamless phase 2/3 trial with a time-to-event outcome as the primary endpoint and a binary response as the surrogate endpoint. The sample design parameters presented in Figure 4 were used for demonstration purpose. In each simulation iteration, we first generate the first stage survival outcome from exponential distribution assuming the median survival time for the control arm in both full population $F$ and target subgroup $S$ as 14 months. The median survival time for the treatment group was calculated under each scenario with corresponding hazard ratio in the full population $\lambda^{(F)}$, and in the subgroup $\lambda^{(S)}$. The surrogate endpoint is assumed to be binary with mean response rate for controls in both the full population ($p_c^{(F)}$) and subgroup ($p_c^{(S)}$) set as 0.2 and that of the treatment groups varied under different scenarios by varying the true risk differences (treatment-control) $\theta^{(F)}$ and $\theta^{(S)}$. In a real trial, we can observe the risk difference from trial data and plug it into a historical model (from prior information) to obtain a predicted log-rank test statistic as described in section 2.2. In simulation, to simplify the data generation steps, we directly draw 'predicted' log-rank test statistics from normal distribution $N\left(\log(\lambda^{(s)}) *\right.$

$$\sqrt{\frac{m^{(s)}}{4}} + \rho * \left(\hat{\theta}^{(s)} + \phi - \theta^{(s)}\right), (1-\rho^2)\right), \text{ for } s = S, F, \text{ where } \rho \text{ denotes the correlation}$$

between the test statistics of the primary and surrogate endpoints under each scenario, and $m^{(s)}$ is the corresponding number of events observed at the interim analysis for each population. Three different levels of correlation between primary of surrogate endpoints, $\rho \in \{-0.3, -0.6, -0.9\}$ were explored. The correlations took negative values numerically because a larger risk difference and a lower log-rank statistic represent better treatment effect. Conceptually, this means a treatment having a higher response rate leads to longer survival time. The parameter, $\phi$, was introduced to examine different scenarios where prediction may be biased. A positive $\phi$ leads to a more optimistic prediction of the primary treatment effect, compared to prior knowledge. On the other hand, a negative $\phi$ represents an effect in the opposite direction. This parameter can be also viewed as misspecification of our prediction model. We set $\phi \in \{0.2, 0, -0.2\}$, corresponding to optimistic, neutral, and pessimistic predictions in the simulation study. To represent a fairly accurately observed surrogate endpoint, we plugged-in the true value for risk difference or draw the observation from a tight neighborhood of its true value.

Under the null hypothesis, we set $\lambda^{(F)} = \lambda^{(S)} = 1$, $p_c^{(F)} = p_c^{(S)} = 0.2$, $\theta^{(F)} = \theta^{(S)} = \phi = 0$. The overall type-I error over different correlations $\rho$ and different weight functions in Equations (2-4) were examined with 100,000 simulation iterations. Under the alternative hypothesis, various scenarios listed in Table 2 were repeated in 10,000 simulated trials. In all simulations, we assumed the prevalence of the targeted subgroup is 0.5, and patients were equally randomized among thetreatment arms. The treatment effect was tested at one-sided $\alpha = 0.025$.

[Table 2]

*3.2 Simulation results*

The percentage of trials falling in different interim decision zones under each scenario is presented in Figure 5. In scenario sets (a) and (c) (left column), where the treatment effect exists and is homogeneous within the full population, the simulation results show that more Favorable, Promizing or Enrichment decisions are made compared to a benchmark design without surrogate information (no SE bar). The probability of being in these zones increases as $\phi$ and the absolute value of $\rho$ increases, with one exception in the set (c) panel $\phi = -0.2$, where there is a slight decrease of these zones when the magnitude of correlation increases. This might be because this panel is a situation where the surrogate information is biased towards the null by setting. In this case, the stronger the correlation is, the trials are more likely to fall into the Unfavorable zone. However, there is still less Futility zone than the benchmark design.

In the scenario sets (b) and (d) (right column), the general trend is similar. With additional surrogate information, the probability of a trial in the Enrichment zone increases compared to the benchmark. This difference is more noticeable in the set (d) where there is more heterogeneity in the treatment effect. The probability of the Enrichment zone could be doubled if there is a high correlation between endpoints and when $\phi = 0.2$. In the set (b), because the hazard ratio 0.7 in the full population provides stronger signal comparing to that in the set (d), more trials tend to go to the Promising zone when $\phi = 0.2$ and correlation increases.

[Figure 5]

Figure 6 shows the overall power of the proposed design under various alternative scenarios. The overall power is defined as the probability to reject both the elementary and intersection null hypotheses based on population selected at the interim analysis as shown in Section 2.4. Under all scenarios, the proposed design has higher overall power compared to the benchmark without surrogate information (black solid line). Within each panel (same value of $\phi$), a higher correlation leads to larger power gain. Specifically, the difference is the greatest in scenario set (d) where the target subgroup has a much lower hazard ratio than its counterpart.

As shown in Table 3, the overall power of the proposed design is higher than that of the benchmark design, where surrogate information is not included in the interim decision making, under different scenarios. The power gain is at the cost of increased event size, and longer study duration in order to observe sufficient number of event to trigger the final analysis. Under scenario set (a), compared to the benchmark design, the proposed design achieves about 4-6% power gain with 9-17 more events on average, while scenarios set (d) reveals a greater power gain (7-14%) for the proposed design with 13-26 more expected events. The trend of power gain is very similar across different $\phi$, demonstrating that the proposed method is robust with misspecification of surrogate information and the historical model.

[Figure 6]

[Table 3]

For the operating characteristic under the alternative hypothesis, we only present the results from the first weight function shown in Equation (2) in this paper. As shown by Li et al. (2021), the performances of the other two weight functions are very comparable.

Under the null hypothesis, our proposed method maintains the overall type I error below the proposed level of $\alpha = 0.025$ (Table 4). Since including futility boundary in a trial leads to a reduction of type I error and there is no efficacy boundary in our design to offset this reduction (Chen et al., 2004; Mehta & Pocock, 2011), the type I error may not be comparable to the designed level. Therefore, another set of simulations of the proposed design without the futility boundary was also conducted and presented in table 4. The overall type I error is about the same level compared to the benchmark design without using surrogate information when estimating CP at the interim analysis. Although it may seem slightly conservative, this phenomenon is consistent with other similar designs shown in literature and is not a big concern for maintaining statistical integrity (Chen et al., 2004; Jenkins et al., 2011; Mehta & Pocock, 2011).

[Table 4]

## 4. Case studies

### *4.1 Oncology case study*

In this section, we will illustrate how to use our proposed design in a seamless phase 2/3 oncology study comparing an investigational regimen A versus a regimen B for patients with multiple myeloma (MM). Multiple myeloma (MM) is a B-cell tumor of malignant plasma cells within the bone marrow.

It remains incurable despite advances in novel therapies with proteasome inhibitors, immunomodulating drugs (IMiD), and stem cell transplant (SCT) therapy. Regimen A is a novel drug combination regimen for multiple myeloma disease. In a phase 1 dose-escalation and safety expansion study, researchers have found exciting safety profiles and a high complete response (CR) rate. Moreover, a much higher than expected CR for high-risk cytogenetic subgroup patients was also observed. Therefore, an adaptive seamless phase 2/3 randomized study with subgroup enrichment feature is planned for selecting the right population for this new drug combination and accelerating the development. To make an informed Go/No-Go decision at the phase 2 stage with limited sample size, the interim decision is planned to utilize all critical efficacy information including both primary endpoint PFS and key secondary endpoint CR. According to previous studies with the same indication, patients with a higher probability of response to CR tends to have better PFS (Harousseau et al., 2010). The numerical correlation between risk difference (treatment-control) and the log-rank test statistic is assumed to be about -0.6 in this case study (Avet-Loiseau et al., 2020; Daniele et al., 2021).

It is expected that a total of 300 patients will be enrolled in this seamless phase 2/3 randomized clinical trial. Assuming an exponential distribution for PFS, an initial minimum of 160 PFS events is required to detect an optimistic hazard ratio of 0.6 with approximately 90% power using a 1-sided log-rank test. Under the same assumptions, a maximum of 240 PFS events is needed to detect a hazard ratio of 0.66. Therefore, the total number of PFS events for the final analysis is set to range from 160 events to 240 events. Assuming 8 patients per month will be enrolled in the phase II stage and 15 patients per month will be enrolled in the phase III stage separately, with an estimated 52 months study duration, the total sample size 300 is estimated to generate 160 to 240 events based on maintaining 90% power to test the primary endpoint PFS at 1-sided alpha level 0.025.

The primary endpoint of this study is PFS and the key secondary endpoint is CR rate. The objective of phase 2 part is to check for futility, to determine the PFS event size, and to select the patient population to be enrolled in the subsequent phase 3 stage. The analysis for phase 2 part will be performed when approximately 40 PFS events are observed in the ITT population. The criteria and all possible interim decisions based on modified CP are shown in Table 5.

[Table 5]

Suppose at the interim analysis, due to immature information from the PFS endpoint, we observe a hazard ratio of 0.98 in the full population and 1.15 in the cytogenetic subgroup. The corresponding log-rank statistics are -0.05 and 0.27 respectively, and the CP for the full population and the cytogenetic subgroup are both below 0.05. By the interim decision rule in Table 5, the trial falls into the Futility zone. Thus, this trial will be stopped early at interim analysis.

However, the decision would be very different if the surrogate endpoint CR is taken into account during interim analysis. Suppose at the interim analysis, the observed risk difference for the full population about 0.19 and subgroup about 0.38. By an assumed historical linear model, the predicted log-rank test statistics are 0.09 and -1.73 for the full population and subgroup respectively. If the weighted function 1 in equation (2) is used, the modified CP is now 0.66 for the cytogenetic subgroup and below 0.05 for the full population. Thus, the interim decision goes to the Enrichment zone, and the corresponding re-estimated event size for the subgroup becomes 168. The trial continues into Phase 3 stage with an expanded event size but limits recruitment to the high-risk cytogenetic subgroup only. After observing 168 events in the subgroup, a final analysis is conducted. With the extended event size and study duration, the final estimate of the hazard ratio is about 0.68 in the subgroup (p=0.006), demonstrating a significant treatment effect in the high cytogenetic subgroup. In summary, our improved design provides an opportunity to rescue a promising drug for the subgroup patients even if the primary endpoint data at interim analysis dis not mature enough.

### *4.2 Vaccine case study*

Our proposed enrichment design is a general framework that applies to various therapeutic areas, e.g. Vaccine (Liu et al., 2021). In this section, we will illustrate a case study in Norovirus (NoV) vaccine trial. Noroviruses have emerged as the most significant cause of non-bacterial gastroenteritis worldwide, with typical symptoms including vomiting, diarrhea, and fever. It is highly infectious and causes acute to severe illness, and poses the most risks to the very young, the elderly, and immunocompromised individuals. Noroviruses represent a significant burden to public health and hence the need for an effective vaccine.

Let's consider a large phase 2/3 seamless Norovirus vaccine efficacy trial, which is a double-blinded, global, randomized, controlled study with a 1:1 randomization ratio. With the proposed design framework showed in this paper, the trial is designed to prove the effective prevention of norovirus-associated acute gastroenteritis (AGE) with population enrichment feature, since earlier trials showed that baseline serostatus (positive or negative) may have a potential impact on the vaccine efficacy. The primary endpoint is vaccine efficacy (VE) defined as $100\%*[1 - (\lambda_v / \lambda_c)]$, where $\lambda_v$ and $\lambda_c$ denote the hazard rates for the vaccine and placebo arms, respectively, based on time to AGE cases. Cox proportional hazards (Cox PH) model is used and the 95% CI for VE is derived from the CI estimation of the hazard ratio obtained from the Cox PH model. The primary efficacy endpoint is considered to be fulfilled if the lower bound of the 95% CI for the VE is above 35%.

For vaccine trials, an interim analysis could be planned with information fraction <50%, to obtain earlier safety and immunogenicity data especially if the target population or region has been changed from previous trials, e.g., from adult to children, or from endemic area to non-endemic area. Due to limited primary endpoint cases collected by the time of the interim analysis, available immunogenicity data, usually antibody titer, could be used as a surrogate endpoint to support better decision making at the interim. Unlike other therapeutic areas, the correlation between vaccine efficacy endpoint and immunogenicity endpoint is established through Correlate of Protection (COP) analyses, usually based on pivotal phase 3 efficacy trial. Therefore, COP may not be available before launching the phase 2/3 trial. One possibility is to employ pre-clinical animal data if available. In this case study, a simplified moderate correlation to be 0.7 is used for illustration purpose.

For a case-driven vaccine efficacy trial, assuming true VE of 70%, a total of 92 cases of first norovirus AGE cases would provide at least 90% power to rule out a treatment effect of ≤ 35% (with a two-sided significance level of 0.05) in the primary endpoint of preventing first occurring AGE due to NoV infection. Approximately 4300 subjects would be needed to assess the primary endpoint assuming a disease rate of 37/1000 and 10% dropout rate. The interim analysis is planned after approximately 26 (30%) of the total AGE cases have been observed, and an increase in both event size and the sample size is allowed by design. Depending on available resources and study timeline, a maximum of 112 AGE cases could be considered, and subject enrollment could increase to a maximum of 5000 to shorten study duration which is lengthened by additional events. Based on these event size parameters, the interim decision rule and corresponding VE levels is shown in Figure 7. The vertical and horizontal axes represent the observed VE and predicted VE with the surrogate endpoint, antibody titer, respectively. The diagonal line denotes the traditional CP decision criteria without considering the surrogate endpoint at interim analysis.

[Figure 7]

Suppose at the interim analysis, due to immature information of AGE events, VE observed are 56% in full population and 65% in the seronegative subgroup, with conditional power 0.34 and 0.62 respectively. Based on the predefined interim decision rule, the trial falls into the Enrichment zone. Thus, the decision is to proceed with only the seronegative subgroup, which is less desirable because it poses challenges for regulatory approval as well as a marketing strategy. However, when taking the surrogate endpoint, antibody titer, into account, the decision could be different. Suppose that at the interim analysis, the observed log (geometric mean titer difference) for the full population is 1.2 and subgroup about 1.5. With an assumed historical model between two endpoints and equation (2), the modified CP is boosted to 0.98 for the seronegative subgroup while 0.77 for the full population. The interim decision has moved to the Promising zone, i.e., continues into Phase 3 stage with the full population, indicating a potential full population vaccine efficacy. After observing 112 events in the full population, the final analysis is conducted. Results show vaccine efficacy is successfully demonstrated with final VE = 62% and lower bound of CI =42%. In summary, in this case study, taking immunogenicity measurement into account for the interim decision leads to the success of the study with the full population.

## 5. Discussion and Conclusion

How to make a reliable interim decision is one of the practical questions for clinical trials employing adaptive designs. In a seamless phase 2/3 subgroup enrichment design, the available information for interim decision-making on whether to enrich the target subgroup is usually very limited, particularly for primary outcomes that require a longer term to observe. On the other hand, a later timed interim analysis allows a longer time to collect information, but there is less room for adaptation for the later stage of the trials. In this paper, we presented a general framework of a seamless phase 2/3 adaptive subgroup enrichment design with consideration of a short-term surrogate endpoint, while also allowing sample size re-estimation. The proposed design demonstrates superior performance, i.e. higher chances of desirable interim decisions and higher overall power, compared to a benchmark design not including information from surrogate endpoints at interim decision making, similar to the one shown in Mehta et al. (2019). Our proposed design makes better informed interim decisions because it takes advantage of all the available information from both the primary and surrogate endpoints at the time of the interim analysis. The overall type I error is controlled by the combination of the CHW test and the closure testing principle.

Some practical considerations for our proposed method include the availability of prior knowledge about the relationship between the primary and surrogate endpoints, as well as the relationship between the targeted subgroup and the general patient population. Our method is most advantageous when the surrogate endpoint is a good predictor of the primary treatment effect. Such a relationship may be easier to identify in some well-studied disease conditions but may be challenging to validate in others. In the simulation study, we demonstrate that our proposed design is robust when there is some overall drift in our estimated historical model from the truth. In practice, it is recommended to use a better understood surrogate endpoint with a higher correlation to the primary endpoint to utilize the full benefit of incorporating modified CP at interim analysis. In our proposed design, we assumed that only one targeted subgroup is of the clinical interest, thus the enrichment feature only focuses on the targeted subgroup and the treatment effect is not tested in its counterparts. There are also some relevant designs considering mutiple subgroups and/or multiple arms, for example, platform trials or other master protocols (Berry, Connor, & Lewis, 2015; Lin & Bunn, 2017; Lin et al., 2019).

The application of the proposed design in two case studies is demonstrated in this paper, one for oncology and the other for vaccine study. This design is also very general and flexible to be extended to many therapeutic areas where there is pressing unmet medical needs, as well as other types of adaptive design where borrowing information from surrogate endpoint can facilitate interim decision making. Although the current framework is developed based on a simple linear relationship between the primary and surrogate endpoints, it can be easily extended to other situations for different disease conditions where more complex models are needed.

The treatment effect estimation in both case studies are based on the conventional approach. However, using a conventional sample mean to estimate treatment effect may induce some biases for adaptive designs. Jennison and Turnbull (1999) and Pallmann (2018), Li, Lin & Lin (2020) summarized some methods for adjusting estimates to reduce or remove bias. Through deriving an bias-reduced or unbiased estimator instead of using the conventional one can be considered to reduce the bias and/or the variance of the treatment effects. The maximum likelihood estimator (MLE) or some likelihood-based method used on adaptive design can be biased and the magnitude of bias will depend on the unknown model parameters. The Shrinkage estimation methods (Carreras and Brannath, 2013 and Bowden, Brannath and Glimm, 2014) could potentially reduce the bias, but not totally eliminate the bias. This bias problem for adaptive design also exists for the confidence interval. Kimani et al., 2020 proposed a novel asymptotically unbiased estimator and a new interval estimator with good coverage probabilities, but it still restricts to certain circumstances. Moreover, Exploring the optimal timing for interim analysis and developing some objective criteria for including different endpoints at interim can also be further explored in future studies.

Table 1. Interim decision rule based on Conditional Power (CP).

| $CP_S, CP_F$ | Zone | Decision |
| --- | --- | --- |
| $CP_F \geq 1 - \beta$ | Favorable | Do not increase sample size |
| $1 - \beta > CP_F \geq \delta^{(F)}$ | Promising | Increase sample size, continue enrollment in full population |
| $CP_F < \delta^{(F)}$ and $CP_S \geq \delta^{(S)}$ | Enrichment | Increase sample size, continue enrollment in subgroup only |
| $CP_F < \delta^{(F)}$ and $CP_S < \delta^{(S)}$ | Unfavorable | Do not increase sample size |

Table 2. Different alternative scenarios explored in simulation study.

| Scenarios | | $\lambda^{(F)}$ | $\lambda^{(S)}$ | $p_c^{(F)}$ | $p_c^{(S)}$ | $\theta^{(F)}$ | $\theta^{(S)}$ | $\phi$ | $\rho$ |
|---|---|---|---|---|---|---|---|---|---|
| a) | 1-3 | 0.6 | 0.6 | 0.2 | 0.2 | 0.4 | 0.4 | 0 | -0.3,-0.6,-0.9 |
| | 4-6 | | | | | | | 0.2 | -0.3,-0.6,-0.9 |
| | 7-9 | | | | | | | -0.2 | -0.3,-0.6,-0.9 |
| b) | 10-12 | 0.7 | 0.6 | 0.2 | 0.2 | 0.2 | 0.3 | 0 | -0.3,-0.6,-0.9 |
| | 13-15 | | | | | | | 0.2 | -0.3,-0.6,-0.9 |
| | 16-18 | | | | | | | -0.2 | -0.3,-0.6,-0.9 |
| c) | 19-21 | 0.7 | 0.7 | 0.2 | 0.2 | 0.3 | 0.3 | 0 | -0.3,-0.6,-0.9 |
| | 22-24 | | | | | | | 0.2 | -0.3,-0.6,-0.9 |
| | 25-27 | | | | | | | -0.2 | -0.3,-0.6,-0.9 |
| d) | 28-30 | 0.8 | 0.6 | 0.2 | 0.2 | 0.2 | 0.4 | 0 | -0.3,-0.6,-0.9 |
| | 31-33 | | | | | | | 0.2 | -0.3,-0.6,-0.9 |
| | 34-36 | | | | | | | -0.2 | -0.3,-0.6,-0.9 |

Table 3 Empirical power, study duration (Dur) in monthes, and total number of events (Evts) under different alternative scenarios, compared to a similar design without considering surrogate endpoint (SE) at interim decision making.

| Scenario | | $\rho=-0.3$ | | | $\rho=-0.6$ | | | $\rho=-0.9$ | | | No SE | | |
|---|---|---|---|---|---|---|---|---|---|---|---|---|---|
| | $\phi$ | Power | Dur | Evts | Power | Dur | Evts | Power | Dur | Evts | Power | Dur | Evts |
| a) | 0.2 | 0.88 | 51 | 176 | 0.87 | 51 | 176 | 0.88 | 50 | 175 | 0.83 | 49 | 167 |
| | 0 | 0.87 | 51 | 176 | 0.89 | 52 | 178 | 0.89 | 52 | 180 | | | |
| | -0.2 | 0.87 | 51 | 177 | 0.87 | 52 | 179 | 0.89 | 54 | 184 | | | |
| b) | 0.2 | 0.70 | 50 | 177 | 0.70 | 51 | 180 | 0.73 | 53 | 187 | 0.63 | 46 | 163 |
| | 0 | 0.69 | 50 | 178 | 0.71 | 51 | 181 | 0.72 | 54 | 189 | | | |
| | -0.2 | 0.70 | 50 | 177 | 0.70 | 51 | 180 | 0.73 | 54 | 189 | | | |
| c) | 0.2 | 0.61 | 49 | 176 | 0.62 | 50 | 178 | 0.64 | 52 | 186 | 0.57 | 45 | 159 |
| | 0 | 0.61 | 49 | 176 | 0.62 | 50 | 179 | 0.64 | 53 | 187 | | | |
| | -0.2 | 0.61 | 49 | 176 | 0.60 | 50 | 177 | 0.63 | 52 | 185 | | | |
| d) | 0.2 | 0.55 | 48 | 174 | 0.57 | 49 | 178 | 0.62 | 52 | 186 | 0.48 | 44 | 160 |
| | 0 | 0.55 | 48 | 173 | 0.57 | 49 | 177 | 0.61 | 52 | 185 | | | |
| | -0.2 | 0.55 | 48 | 174 | 0.56 | 49 | 176 | 0.60 | 51 | 183 | | | |

Table 4. Empirical Type I error with different correlation assumption and weight functions $CP_s^{(1)}(\cdot)$, $CP_s^{(2)}(\cdot)$, $CP_s^{(3)}(\cdot)$ corresponding to Equation (2)-(4) for estimating MCP under the null hypothesis.

|  | Type I error (no Futility zone) | | | Type I error (with Futility zone) | | |
| --- | --- | --- | --- | --- | --- | --- |
| No SE | 0.022 | | | 0.019 | | |
| Correlation | $CP_s^{(1)}(\cdot)$ | $CP_s^{(2)}(\cdot)$ | $CP_s^{(3)}(\cdot)$ | $CP_s^{(1)}(\cdot)$ | $CP_s^{(2)}(\cdot)$ | $CP_s^{(3)}(\cdot)$ |
| $\rho = -0.3$ | 0.020 | 0.022 | 0.020 | 0.016 | 0.018 | 0.014 |
| $\rho = -0.6$ | 0.020 | 0.022 | 0.020 | 0.016 | 0.018 | 0.013 |
| $\rho = -0.9$ | 0.020 | 0.020 | 0.019 | 0.018 | 0.016 | 0.010 |

Table 5. Interim decisions and criteria for the oncology case study.

| Interim criteria | Interim Decision | Phase 3 population | Event size | Final analysis |
|---|---|---|---|---|
| $CP_F \geq 0.9$ | Favorable | Full population | 160 | Testing treatment effect in full population |
| $0.9 > CP_F \geq 0.4$ | Promising | Full population | Adaptive, (160,224) | Testing treatment effect in full population |
| $0.4 > CP_F \geq 0.05$ and $CP_S \geq 0.5$ | Enrichment | Cytogenetic subgroup | Adaptive, (160,224) | Testing treatment effect in the cytogenetic subgroup |
| $(0.4 > CP_F \geq 0.05$ and $0.5 > CP_S)$; or $(0.05 > CP_F$ and $0.5 > CP_S \geq 0.05)$ | Unfavorable | Full population | 160 | Testing treatment effect in full population |
| $CP_F < 0.05$ and $CP_S < 0.05$ | Futility | - | - | - |

Figure 1. A schematic plot of different adaptive interim decisions in a seamless phase 2/3 design.

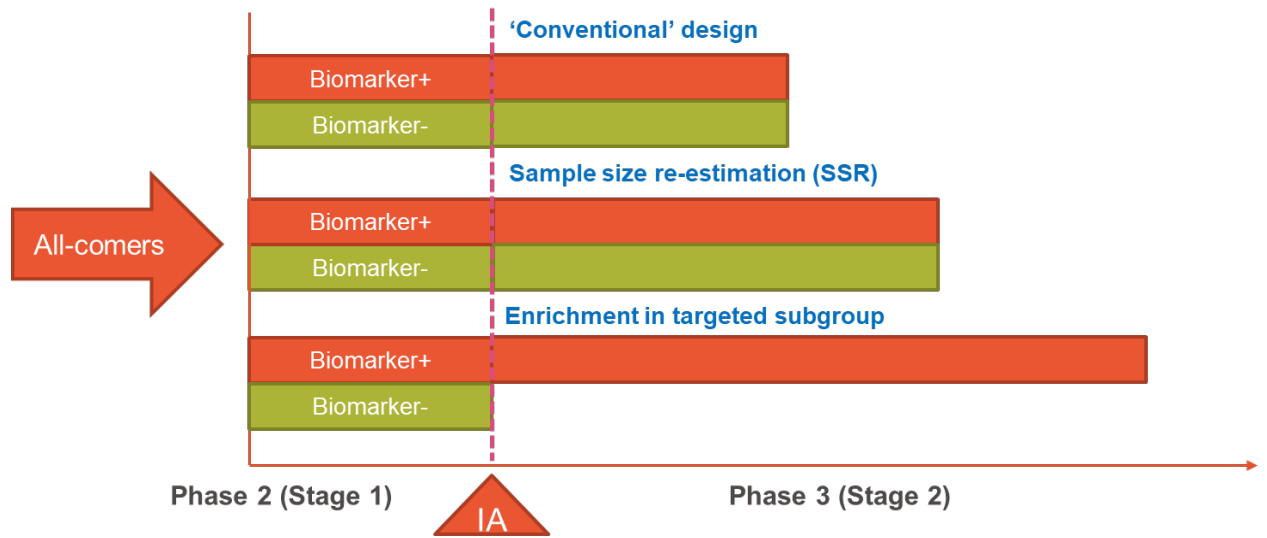

Figure 2. A conceptual plot for long-term survival endpoint and short-term surrogate endpoint observed at interim analysis.

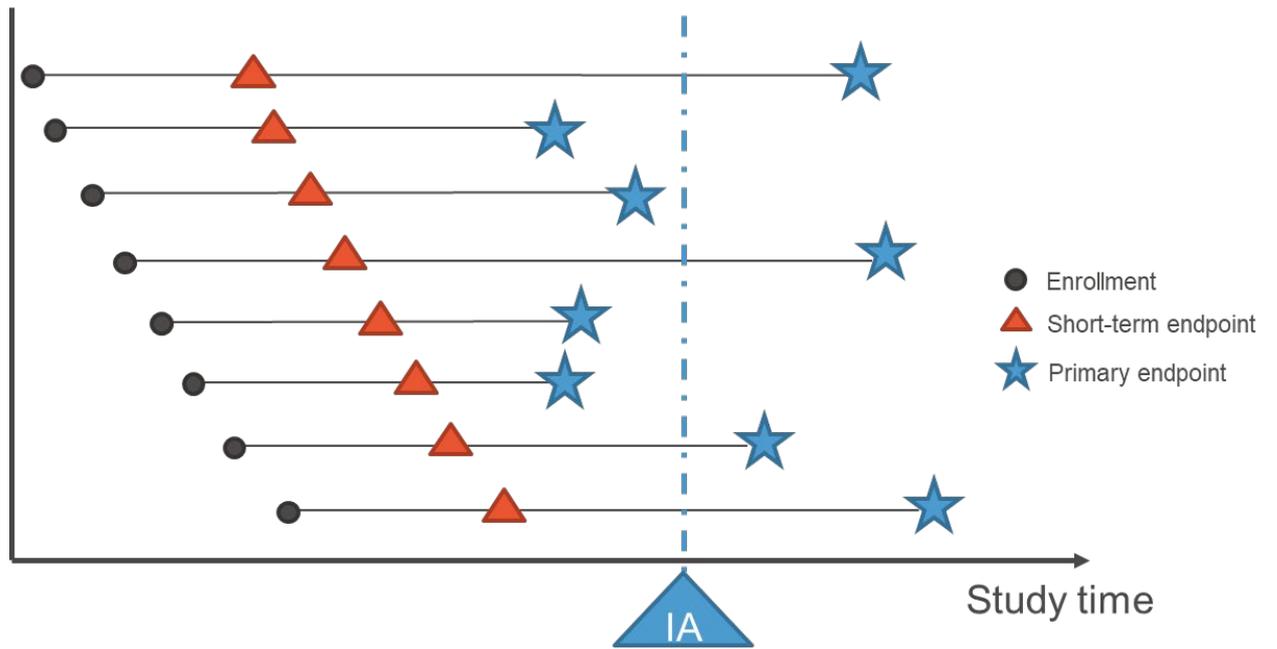

Figure 3. Interim decision rules.

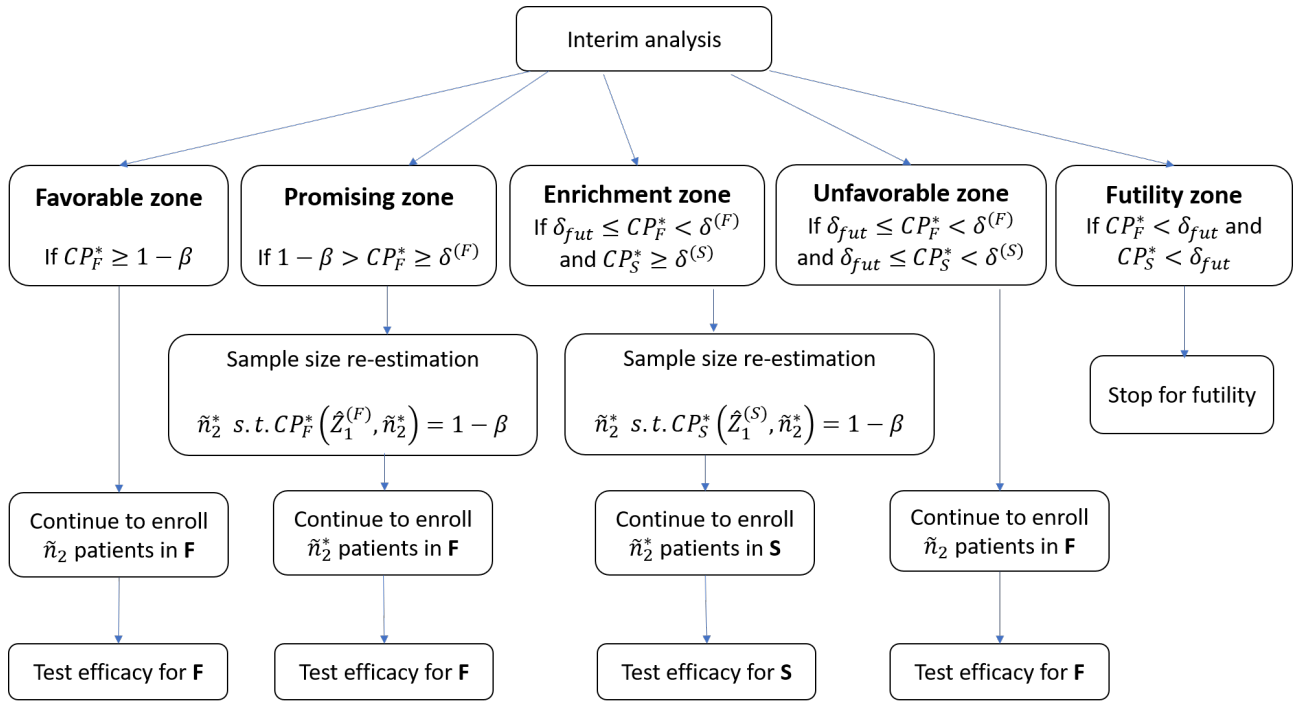

Figure 4. Example design. (a) Sample size parameters; (b) Interim decision cutoffs.

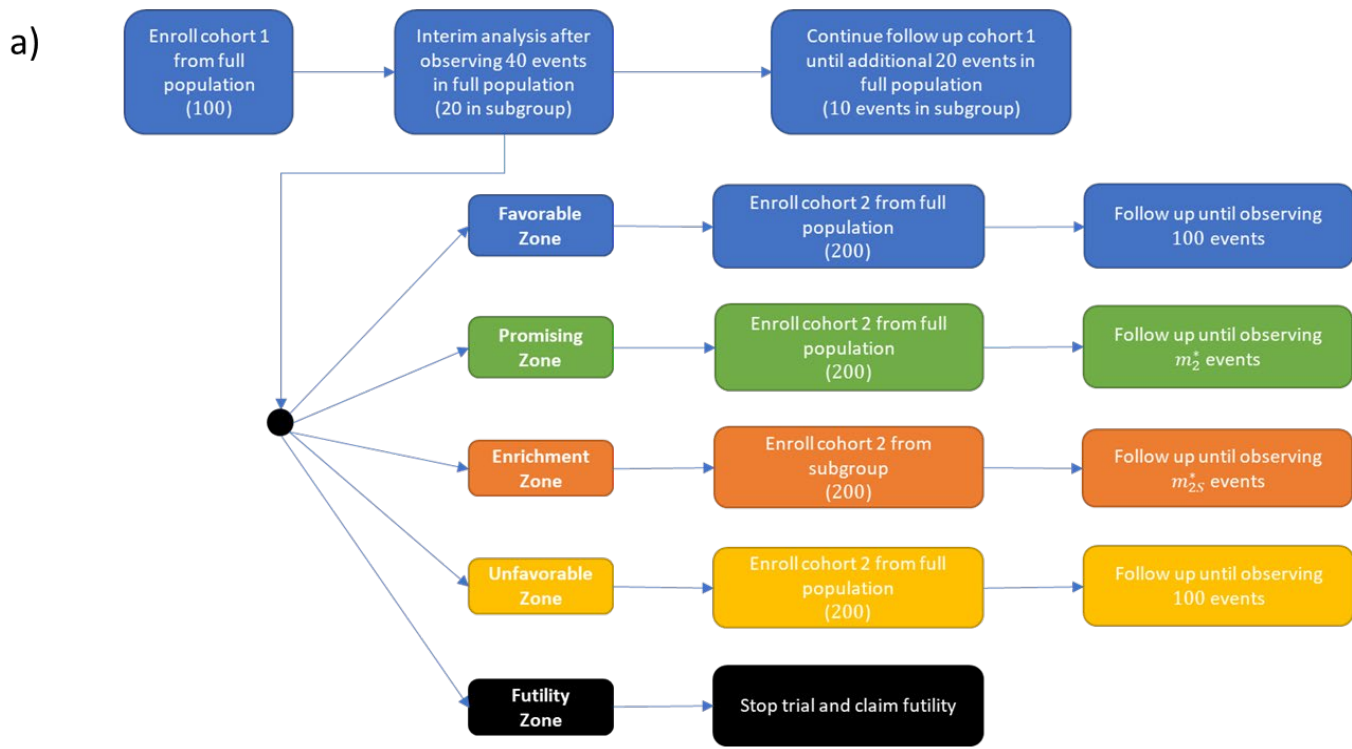

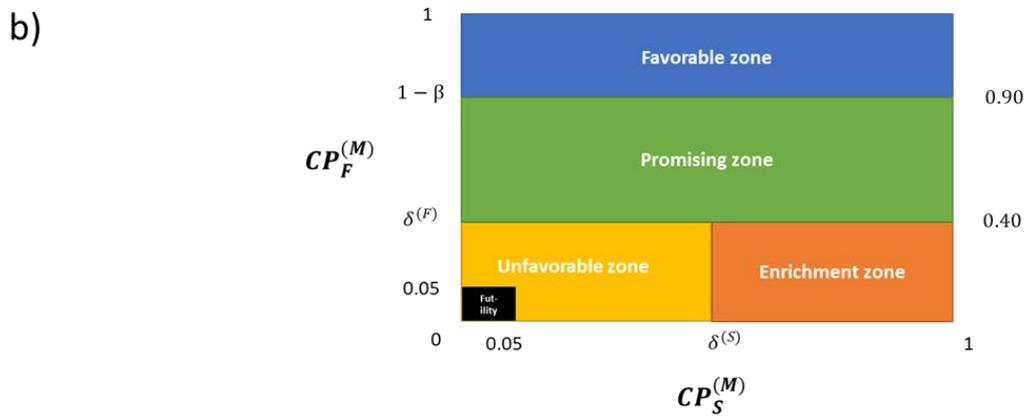

Figure 5. The percentage of trials falling in each interim decision zones under different scenarios paneled by value of the over-expectation offset φ, compared to a benchmark design without surrogate endpoint(SE). The interim decition is based on MCP calculated by Equation (2).

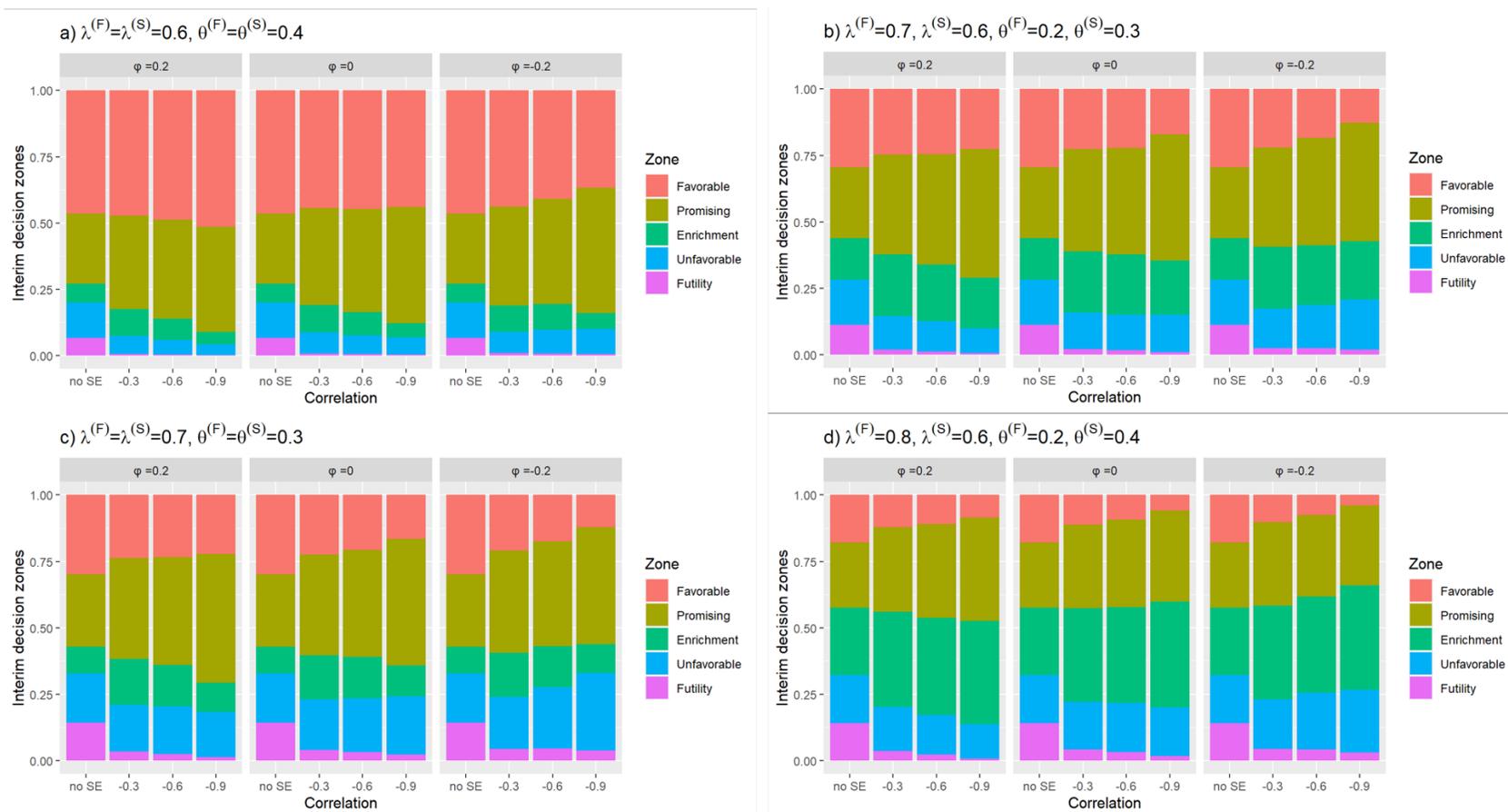

Figure 6. Overall power under different scenarios paneled by value of the over-expectation offset ϕ. The solid black line benchmarks the overall power for a design without surrogate endpoint (SE) in interim decision making. The interim decition is based on MCP calculated by Equation (2).

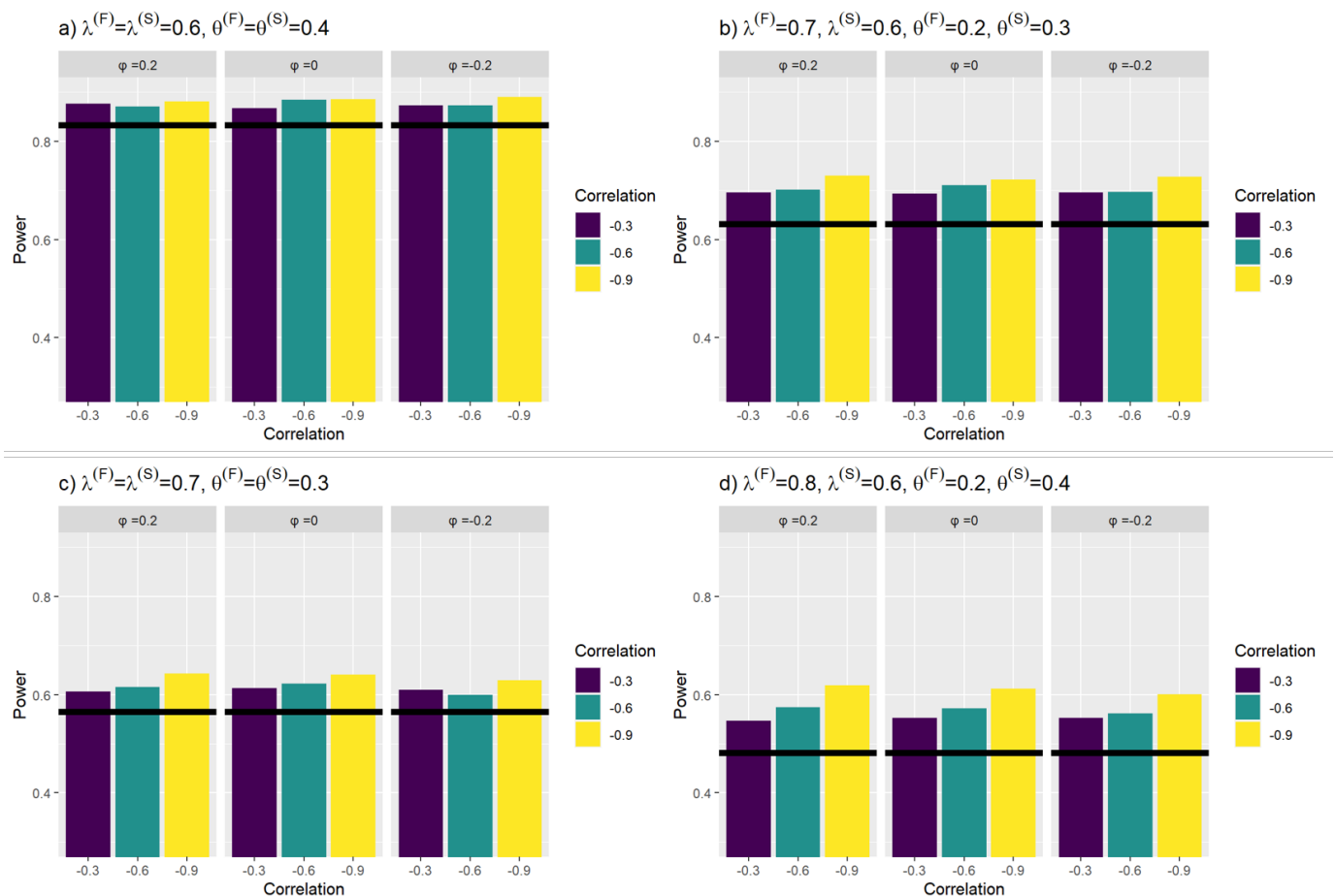

Figure 7. Interim decision rules for vaccine case study.

a) Full population

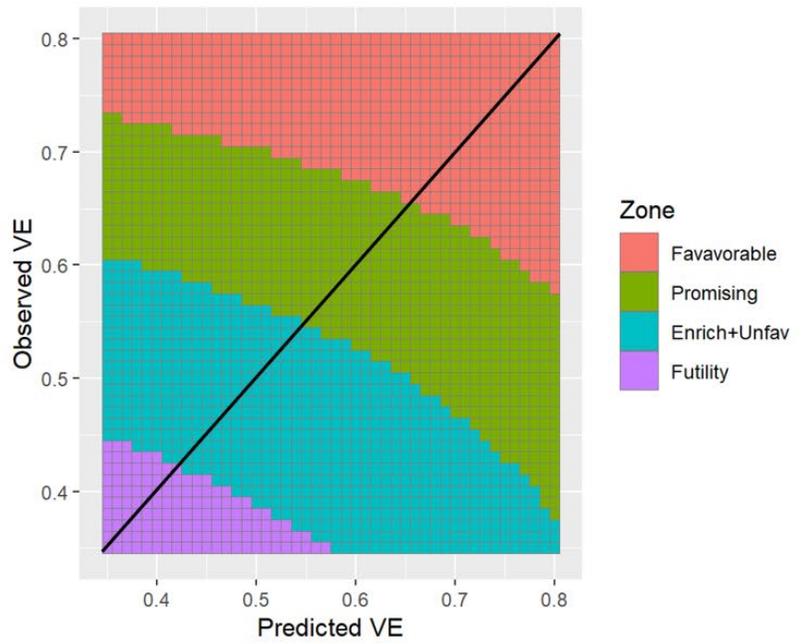

b) Suronegative subgroup

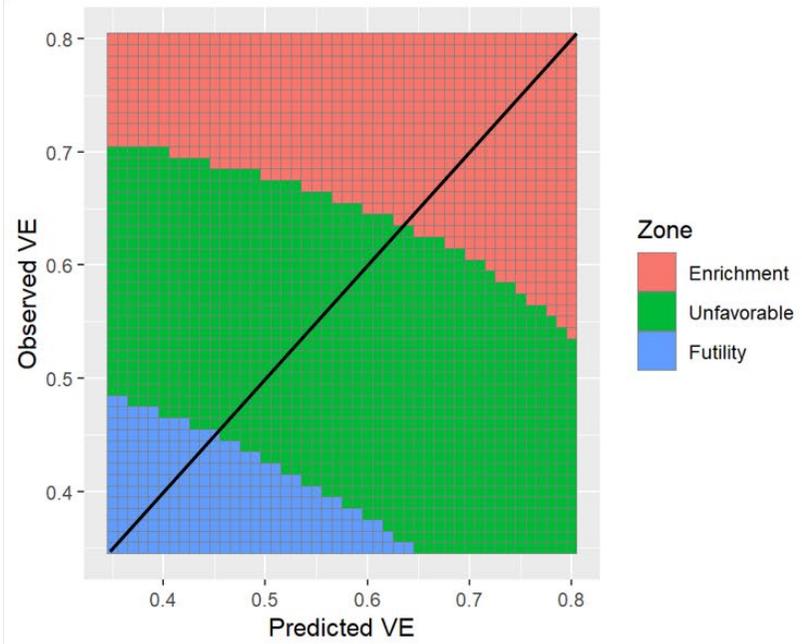